\documentstyle[psfig]{mn}

\newif\ifAMStwofonts

\def\simgt{\hbox{\rlap{\raise 0.425ex\hbox{$>$}}\lower 0.65ex\hbox{$\sim$}}}
\def\simlt{\hbox{\rlap{\raise 0.425ex\hbox{$<$}}\lower 0.65ex\hbox{$\sim$}}}
\def\h1{$h^{-1}$}

\ifoldfss
  \ifCUPmtlplainloaded \else
    \NewTextAlphabet{textbfit} {cmbxti10} {}
    \NewTextAlphabet{textbfss} {cmssbx10} {}
    \NewMathAlphabet{mathbfit} {cmbxti10} {} 
    \NewMathAlphabet{mathbfss} {cmssbx10} {} 
  \fi
  \ifAMStwofonts
    \ifCUPmtlplainloaded \else
      \NewSymbolFont{upmath} {eurm10}
      \NewSymbolFont{AMSa} {msam10}
      \NewMathSymbol{\upi}     {0}{upmath}{19}
      \NewMathSymbol{\umu}     {0}{upmath}{16}
      \NewMathSymbol{\upartial}{0}{upmath}{40}
      \NewMathSymbol{\leqslant}{3}{AMSa}{36}
      \NewMathSymbol{\geqslant}{3}{AMSa}{3E}

    \fi
  \fi
\fi 

\ifnfssone
  \newmathalphabet{\mathit}
  \addtoversion{normal}{\mathit}{cmr}{m}{it}
  \addtoversion{bold}{\mathit}{cmr}{bx}{it}
  \newmathalphabet{\mathbfit} 
  \addtoversion{normal}{\mathbfit}{cmr}{bx}{it}
  \addtoversion{bold}{\mathbfit}{cmr}{bx}{it}
  \newmathalphabet{\mathbfss} 
  \addtoversion{normal}{\mathbfss}{cmss}{bx}{n}
  \addtoversion{bold}{\mathbfss}{cmss}{bx}{n}
  \ifAMStwofonts
    \ifCUPmtlplainloaded \else
      \UseAMStwoboldmath
      \makeatletter
      \new@mathgroup\upmath@group
      \define@mathgroup\mv@normal\upmath@group{eur}{m}{n}
      \define@mathgroup\mv@bold\upmath@group{eur}{b}{n}
      \edef\UPM{\hexnumber\upmath@group}
      \new@mathgroup\amsa@group
      \define@mathgroup\mv@normal\amsa@group{msa}{m}{n}
      \define@mathgroup\mv@bold\amsa@group{msa}{m}{n}
      \edef\AMSa{\hexnumber\amsa@group}
      \makeatother
      \mathchardef\upi="0\UPM19
      \mathchardef\umu="0\UPM16
      \mathchardef\upartial="0\UPM40
      \mathchardef\leqslant="3\AMSa36
      \mathchardef\geqslant="3\AMSa3E
    \fi
  \fi
\fi 
\ifnfsstwo
  \DeclareMathAlphabet{\mathbfit}{OT1}{cmr}{bx}{it}
  \SetMathAlphabet\mathbfit{bold}{OT1}{cmr}{bx}{it}
  \DeclareMathAlphabet{\mathbfss}{OT1}{cmss}{bx}{n}
  \SetMathAlphabet\mathbfss{bold}{OT1}{cmss}{bx}{n}
  \ifAMStwofonts
    \ifCUPmtlplainloaded \else
      \DeclareSymbolFont{UPM}{U}{eur}{m}{n}
      \SetSymbolFont{UPM}{bold}{U}{eur}{b}{n}
      \DeclareSymbolFont{AMSa}{U}{msa}{m}{n}
      \DeclareMathSymbol{\upi}{0}{UPM}{"19}
      \DeclareMathSymbol{\umu}{0}{UPM}{"16}
      \DeclareMathSymbol{\upartial}{0}{UPM}{"40}
      \DeclareMathSymbol{\leqslant}{3}{AMSa}{"36}
      \DeclareMathSymbol{\geqslant}{3}{AMSa}{"3E}
    \fi
  \fi
\fi 

\ifCUPmtlplainloaded \else
  \ifAMStwofonts \else 
    \def\upi{\pi}
    \def\umu{\mu}
    \def\upartial{\partial}
  \fi
\fi

\title{Undistorted Lensed Images in Galaxy Clusters}
\author[L. L. R. Williams and G. F. Lewis]
       {L. L. R. Williams$^{1}$\thanks{Email: {\bf \tt llrw@ast.cam.ac.uk}} 
and G. F. Lewis$^{1,2}$\thanks{Email: {\bf \tt gfl@brenda.ess.sunysb.edu}} \\
${^1}$Institute of Astronomy, Madingley Road, Cambridge, CB3 0HA, U.K. \\
${^2}$Astronomy Group, Dept. of Earth and Space Sciences, 
SUNY at Stony Brook, NY 11794-2100, U.S.A. }
\date{Accepted to MNRAS}
\pubyear{1997}

\begin{document}

\maketitle

\newcommand{\fmmm}[1]{\mbox{$#1$}}
\newcommand{\scnd}{\mbox{\fmmm{''}\hskip-0.3em .}}
\newcommand{\scnp}{\mbox{\fmmm{''}}}

\begin{abstract}
To date, the study of high-magnification gravitational lensing effects 
of galaxy clusters has focused upon the grossly distorted, luminous
arc-like features formed in massive, centrally condensed clusters.  We
investigate the formation of a different type of image, highly magnified
yet undistorted, in two widely employed cluster mass density profiles;
an isothermal sphere with a core, and a universal dark matter halo profile
derived from numerical simulations of Navarro et al. We examine the 
properties of images of extended sources produced by these two clusters
profiles, paying particular attention to the undistorted images. Using
simple assumptions about the source and lens population, we estimate
the relative frequency of occurrence of highly magnified, undistorted 
images and the more commonly known giant arcs.
\end{abstract}


\section{Introduction and motivation}\label{intro}

Gravitational lensing by rich galaxy clusters produces highly
magnified, grossly distorted images of background galaxies (Fort \&
Mellier 1994, Schindler et al. 1995, Lavery 1996). The mechanism of
production of these ``giant arcs'' is well understood; in fact,
the observed arcs are commonly used to determine the mass
distribution in the lensing clusters (Lavery \& Henry 1988,
Fort \& Mellier 1994, Kneib et al. 1995). Despite their low surface
brightness, the giant arcs are highly 
visible and easily detected due to their unique morphology. 

However, it is possible that galaxy clusters are
capable of producing highly magnified yet {\it undistorted} images of
background sources. Whether such images are produced in reasonable numbers
depends on the cluster mass profiles: steep density gradients produce
thin arc-like images (Hammer 1991, Wu \& Hammer 1993), 
whereas flat central cores can easily give rise to undistorted images. 
Because of their regular morphology these latter images
would not be as prominent as arcs, and so would not be readily
recognized and studied.

Undistorted magnified images are the subject of the present paper.
Our aim is two-fold. First, we ask whether realistic
cluster profiles can produce highly magnified undistorted images, or
HMUs, and if so, under what conditions. We consider two types of
clusters: an isothermal model with a core, and a universal dark matter
cluster model derived from the numerical simulations of Navarro, Frenk,
\& White (1995, 1996). The major
difference between these two profiles occurs in their central regions:
isothermal models have a flat density core, whereas those drawn from 
numerical simulations are characterized by gradually flattening 
logarithmic slope, with decreasing radius, and are singular at the 
centre. We examine the image
properties produced by these two profiles, paying particular attention
to HMUs. Second, we use simple assumptions about cluster and source
properties to estimate how common such HMU images might be. As a way
to avoid some parameter dependencies, we do not calculate the actual
frequency of occurrence of such images, instead we estimate the {\it
ratio} of the frequency of HMUs to giant arcs for the two different 
cluster mass profiles. In the present paper we calculate this ratio
assuming circularly symmetric clusters, whereas real clusters exhibit
substructure which can greatly increase the clusters' cross-section
for generating arcs (Bartelmann et al. 1995). However our results 
can be readily converted to apply to substructured clusters, as 
outlined in the Conclusions.

Our results indicate that the statistics of HMUs are quite dependent on 
the overall form of the lensing potential, suggesting a diagnostic tool
of cluster cores. Such a probe of clusters centres is welcome as current 
observational techniques have yet to resolve the matter distribution 
in the inner regions of clusters. Dynamical methods are generally 
inadequate in addressing the issue because of their reliance on 
assumptions (cluster is relaxed, mass follows light), as well as the 
small numbers of galaxy radial velocities (eg. Sadat 1997). X-ray methods
are more promising because the emitting gas is in hydrostatic equilibrium
with the cluster gravitational potential. X-ray  derived cores
tend to be large; based on the whole sample of the EMSS clusters,
Henry et al. (1992) estimate that the average cluster core radius is 
125\h1~kpc. These results are in apparent contradiction with 
strong lensing analysis of clusters. Le F\`{e}vre et al. (1994), 
based on a sample of 16 brightest high-$z$ EMSS clusters, conclude that 
arc statistics are incompatible with X-ray core radii, and much smaller 
cores are required. However, Waxman \& Miralda-Escud\'e (1995), and 
Miralda-Escud\'e \& Babul (1995) pointed out that X-ray and lensing 
observations can be reconciled, and both are compatible with a singular 
dark matter potential like the universal dark matter profile, since 
multi-phase cooling flow gas in this type of a potential tends to be 
isothermal, and naturally produces cores. In any case, X-ray observations 
do not constrain cluster dark matter cores. That leaves us with lensing.
Lensing is favored among other techniques because it directly probes the
distribution of all mass regardless of its baryonic/non-baryonic
nature, dynamical and physical state. Strong lensing observations place
a tight upper limit on cluster cores;
however it is possible that lensing clusters with giant arcs are a
biased sample of clusters, because steeper-than-average density
profiles are more likely to produce thin long arcs. Consequently,
cores of $\sim$50\h1~kpc, or even larger, may still prove to be a common 
feature of clusters.
In fact, there is tentative observational evidence for the existence 
of cluster cores: {\it (i)} Cl0024+1654 is a spectacular cluster-lens 
with five HST-resolved images of a high-$z$ galaxy. The presence of the 
central demagnified image $E$ requires the cluster to have a finite core
(Colley et al. 1996, Smail et al. 1996); 
{\it (ii)} An extended highly luminous $z=2.7$
galaxy was recently discovered $6\arcsec$ from the center of an
intervening cluster (Yee et al. 1996).  Even though its true `lensing
status' is still uncertain, its properties can be explained if the
lensing cluster had a core (Williams \& Lewis 1996, 
but see Seitz et al. 1997).
Thus the existence and sizes of flat cluster cores is still a matter 
of debate.

If found, HMU images themselves would prove useful for the detailed
study of galaxies at high redshift. Many highly distorted arcs have
already been studied to deduce the properties of the high-$z$
sources. For example, their star formation rate can be derived
because spectra of magnified sources can be obtained with a reasonable
telescope integration time~\cite{ebbels1996}. High-resolution HST 
observations of lensed galaxies are used to infer galaxy morphologies
(Smail et al. 1996). However, the distorted
appearance of arcs, coupled with the complicated nature of the lens
model, implies that lensing inversions, which are needed in order to
derive the kinematic and structural properties of the source galaxy,
are difficult. Ideally, one would like to observe the overall
simpler case of highly magnified undistorted images.

\section{Cluster mass profile models}\label{models}

We consider two circularly symmetric cluster mass profiles: 
isothermal sphere with a core (ISC) and the Navarro et al. (1996) 
universal dark matter profile (NFW).

The 2D projected mass density profile of ISC model is given by
\begin{equation}
\Sigma(r)=\Sigma_0{{1+p(r/r_c)^2}\over{[1+(r/r_c)^2]^{2-p}}}.
\label{mass_dist1}
\end{equation}
We adopt $p=0.5$ to obtain an isothermal sphere at large radii; the
ISC model. However any value between 0 and 0.5 results in a realistic
mass model; a $p=0$ represents a Plummer model [see Schneider et al. (1992)
a detailed description of the profile]. In Equation~\ref{mass_dist1}, $r_c$
is the core radius, and $\Sigma_0$ is the central surface mass density.

\def\la{\mathrel{\mathchoice {\vcenter{\offinterlineskip\halign{\hfil
$\displaystyle##$\hfil\cr<\cr\sim\cr}}}
{\vcenter{\offinterlineskip\halign{\hfil$\textstyle##$\hfil\cr
<\cr\sim\cr}}}
{\vcenter{\offinterlineskip\halign{\hfil$\scriptstyle##$\hfil\cr
<\cr\sim\cr}}}
{\vcenter{\offinterlineskip\halign{\hfil$\scriptscriptstyle##$\hfil\cr
<\cr\sim\cr}}}}}
\def\ga{\mathrel{\mathchoice {\vcenter{\offinterlineskip\halign{\hfil
$\displaystyle##$\hfil\cr>\cr\sim\cr}}}
{\vcenter{\offinterlineskip\halign{\hfil$\textstyle##$\hfil\cr
>\cr\sim\cr}}}
{\vcenter{\offinterlineskip\halign{\hfil$\scriptstyle##$\hfil\cr
>\cr\sim\cr}}}
{\vcenter{\offinterlineskip\halign{\hfil$\scriptscriptstyle##$\hfil\cr
>\cr\sim\cr}}}}}

The 3D mass density profile of the NFW model is,
\begin{equation}
\rho(r)={{\rho_s~r_s}\over{r(1+r/r_s)^2}},
\label{mass_dist2}
\end{equation}
where $r_s$ is the scale radius, and $\rho_s$ is the mass density at
$r\approx 0.466~r_s$. This model is derived from N-body simulations of
large scale structure and is found to be applicable on scales of
$3\times10^{11} \la M_{200}/M_\odot \la 3\times10^{15}$, where
$M_{200}$ is the mass within a radius $r_{200}$. Within this radius
the mean overdensity is a factor of 200.

We use normalised lengths, $x=r/r_c$ and $x=r/r_s$, respectively, and 
express the surface mass density in terms of the critical surface 
density for lensing,
\begin{equation}
\kappa(x)=\frac{\Sigma(x)}{\Sigma_{crit}},{\hskip0.5in}
\Sigma_{crit}=\frac{c^2}
{4 \pi G}\frac{D_{os}}{D_{ol}D_{ls}}, 
\label{critical}
\end{equation}
where $D_{ij}$ are the relative angular diameter distances between the
observer $(o)$, lens $(l)$ and source $(s)$.  

For ISC model, the normalized surface mass density is
\begin{equation}
\kappa(x)=\kappa_0{{1+px^2}\over{[1+x^2]^{2-p}}}.
\label{mass_dist_proj1}
\end{equation}
For NFW model, the corresponding equation was calculated by Bartelmann
(1996), who showed that NFW model is compatible with the existence of
radial arcs in clusters. We reproduce the relevant equations here for 
completeness;
\begin{equation}
\kappa(x)=2\kappa_s{f(x)\over{x^2-1}},{\hskip 0.1in}{\rm where}{\hskip 0.1in}
\kappa_s=\rho_s r_s/\Sigma_{crit},{\hskip 0.2in}
\label{mass_dist_proj2a}
\end{equation}
\begin{equation}
{\rm and}{\hskip 0.1in}
  f(x) = \cases{
  1-{2\over\sqrt{x^2-1}}{\rm arctan}\sqrt{x-1\over x+1} & $(x>1)$ \cr
  1-{2\over\sqrt{1-x^2}}{\rm arctanh}\sqrt{1-x\over 1+x} & $(x<1)$ \cr
  0 & $(x=1)$ \cr
  }
\label{mass_dist_proj2b}
\end{equation}

For a circularly symmetric lens, the lens equation, which is the relation
between the source position $y$, the image position $x$, and the
deflection angle, is given by
\begin{equation}
y=x-{\frac{{\int_0^x}2x^\prime \kappa(x^\prime) dx^\prime}{x}}
=x-{\frac{m(x)}{x}},
\label{lens_eq}
\end{equation}
where $m(x)$ is the dimensionless mass interior to $x$, in units of 
$(c^2/4G)(D_{os}D_{ol}/D_{ls})$. 

A lensed image is distorted both radially and tangentially with
respect to the centre of the lensing potential. The tangential distortion 
is the ratio of the tangential size of the image to that of the source; 
for a sufficiently small source it is given by the ratio of their 
respective distances from lens centre,~$x/y$. 
Similarly, the radial distortion is $dx/dy$, if the source's
radial extent is $dy$. Since
magnification is just the ratio of the size of the image to that of
the source, it is given by
\begin{equation}
\mu=\Big\vert {{dx}\over{dy}}{{x}\over{y}}\Big\vert.
\label{amplif}
\end{equation}
The length-to-width ratio of an image whose source is circular is,
\begin{equation}
L/W=\Big\vert {{dy}\over{dx}}{{x}\over{y}}\Big\vert,
\label{LW}
\end{equation}
and is the most commonly used measure of image distortion. Note that
an isothermal density profile, $\kappa(x)\propto x^{-1}$,
implies that $dy/dx=1$ (Equation~\ref{lens_eq}), 
i.e. images suffer no radial (de)magnification.
Steeper profiles always result in radially demagnified images, while 
shallower profiles usually produce radially magnified images.

Using Equations~\ref{mass_dist_proj1}-\ref{LW} we can calculate all the
image properties needed in this paper.

\section{Image Properties}\label{images}

\subsection{ISC model}\label{ISCmodel}

\begin{figure}
\vbox{
\centerline{
\psfig{figure=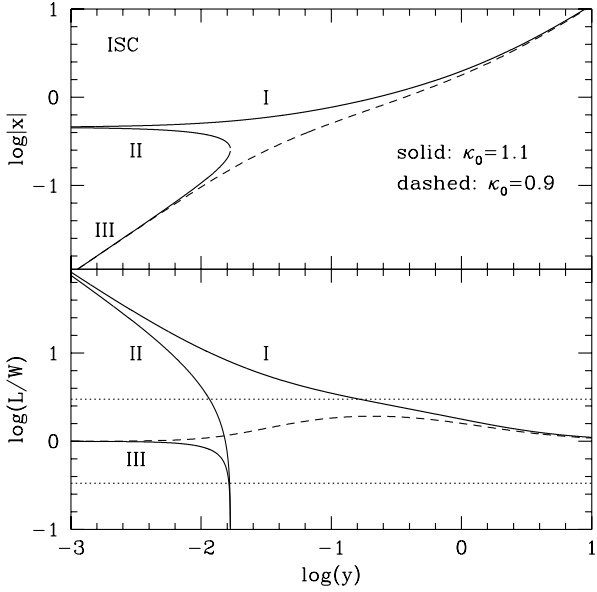,width=3.25in,angle=0}
}
\caption[]{Image position, $\vert x\vert$, (top panel) and
length-to-width ratio, (bottom panel) vs. source impact parameter,
$y$, for two ISC lenses: $\kappa_0=1.1$ (solid lines) and
$\kappa_0=0.9$ (dashed line) of the ISC cluster model.  Roman numerals
indicate image type.  The dotted lines in the bottom panel bracket the
`undistorted' image region, see Section~\ref{criteria}.
\label{LWvsx1}
}
}
\end{figure}

The top panel of Figure~\ref{LWvsx1} shows the relation between the
image and source positions, for two values of $\kappa_0$, of an ISC
model.  Supercritical clusters, represented here by $\kappa_0=1.1$
case (solid lines), produce three images if the source impact
parameter is smaller than the radial caustic, $y_r$ [see Eqn (8.42) of
Schneider et al. (1992)]. 
Subcritical clusters, such as $\kappa_0=0.9$ case (dashed line) always
produce one image, and the $x$-$y$ relation is one-to-one and
monotonically increasing.

\begin{figure}
\vbox{
\centerline{
\psfig{figure=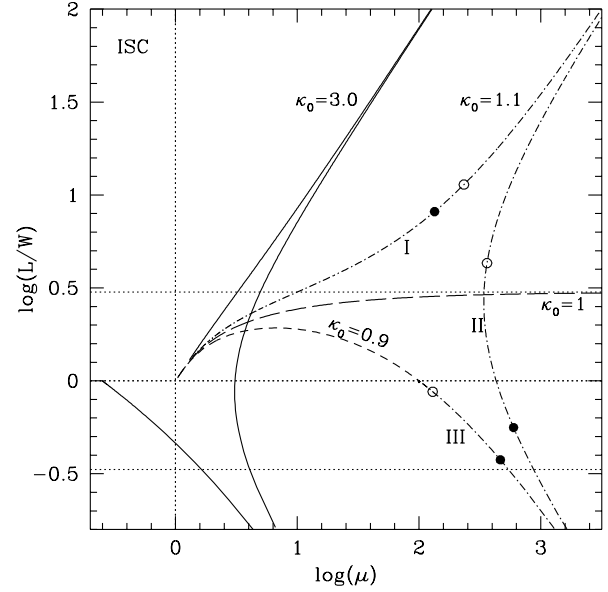,width=3.25in,angle=0}
}
\caption[]{Magnification vs. distortion of images formed by an isothermal
cluster with a core, ISC, represented by Equation~\ref{mass_dist1}.
Supercritical clusters ($\kappa_0=3.0$, solid lines; $\kappa_0=1.1$,
dot-dash lines) have three image branches each. Just-critical 
($\kappa_0=1.0$, long-dash line), and subcritical 
($\kappa_0=0.9$, short-dash line) clusters always have one image.
The images are labeled as in Figure~\ref{LWvsx1}.
Horizontal dotted lines at $L/W=3$ and ${1\over 3}$ mark 
the region of undistorted images. The three empty (solid) circles are 
images of a source at $y=0.010$ ($y=0.016$). Notice that for the primary 
images of supercritical lenses, $\log{(\mu)}$ is proportional to
$\log{(L/W)}$. Just subcritical lenses, on the other hand, produce 
images of high magnification and small distortion.
\label{LWvsAmp1}
}
}
\end{figure}

Figure~\ref{LWvsAmp1} shows the relation between image magnification
and distortion. Two supercritical cases are shown: solid lines are the
three images of a $\kappa_0$=3.0 lens, and dot-dash lines are for a
$\kappa_0$=1.1 lens. The long-dash line is the single image of a
critical lens, $\kappa_0$=1.0, while the short-dash line is the single
image of a $\kappa_0$=0.9 lens.  Each lens with $\kappa_0>1$ has three
branches corresponding to three images. The primary image, which is
formed at the minimum of the lensing potential [see Schneider
(1985) and Blandford \& Narayan (1986)], is labeled I.  This image
appears on the same side of the lens centre as the unlensed source,
and is tangentially extended into an arc, with $L/W>1$. It is always
magnified with respect to the source. The image
formed at the saddle-point of the lensing potential,
sometimes called the `counter arc', possesses reversed parity and is
labeled II. The central image, labeled III, is formed at the maximum
of the lensing potential, and is radially extended, i.e. has
$L/W<1$.  Images II and III are on the side of the lens opposite to
the location of the source.  Note that for very centrally condensed
lenses, i.e. those with $\kappa_0>2$, image III can be
demagnified, $\mu<1$, if the source is sufficiently close to the lens
centre.

The `hidden' parameter in this plot is the source position, $y$. To
illustrate its influence on the location of the images, we have 
plotted the three images of a source at $y=0.010$
(empty circles), and the three images of a source at $y=0.016$ (solid
dots), for a $\kappa_0$=1.1 lens. As the source approaches the lens
centre in projection, images I and II get very elongated and tend to 
merge along the tangential critical line; while image III moves close 
towards the lens centre. 

Notice that the single image branch of a
$\kappa_0<1$ lens joins to and continues as the branch of the central
image, labeled III, of the corresponding $2-\kappa_0$ lens. As an
example, the image of a $\kappa_0=0.9$ lens (short-dash line)
continues as the central image branch of a
$\kappa_0=1.1$ lens. This can be shown as follows.  For image
position $x$ very close to the lens centre image magnification,
$\mu\approx(1-\kappa_0)^{-2}$, has the same numerical value for a
$\kappa_0$ as well as a $2-\kappa_0$ lens.  The distortion $L/W$ of
the image located at the centre is 1, from symmetry.
Therefore, for small $x$, the branches of central images of a
$\kappa_0$ and a $2-\kappa_0$ lens meet at the same point in the
log$(L/W)$ vs. log($\mu$) diagram. The slope of both the branches at 
this point can be shown to be $d\log{(L/W)}/d\log{(\mu)}=-{1\over 2}$, 
independent of $\kappa_0$, and $p$; therefore these two branches are
continuous.

The most visible feature in a lensing cluster is the primary arc,
since it is always well displaced from the cluster centre, is highly 
elongated and always magnified. For this image, the magnification is 
an increasing function of distortion. This is why high magnification 
is associated with high distortion of lensed images in galaxy clusters.
However high magnification need not always imply high distortion.

It is apparent from Figure~\ref{LWvsAmp1} that subcritical clusters,
represented here by a $\kappa_0=0.9$ case, can produce highly
magnified undistorted images\footnote{We give our definition of 
magnified and undistorted later in Section~\ref{criteria}, but for now 
we assume that such images have magnification $\simgt$10, and aspect 
ratio $\simlt$ 3.}. For example, if the source is located close to the
centre of a $\kappa_0=0.9$ lens, its magnification is $\sim$ 100,
while its distortion is negligible. In fact, the
largest $L/W$ ratio attained by an image of any subcritical lens
is not greater than $L/W=3$, as we now show.

The length-to-width ratio of any image in the ISC 
cluster model can be derived from Equations~\ref{LW}, \ref{lens_eq},
and \ref{mass_dist_proj1};
\begin{equation}
L/W={{1-\kappa_0(1+x^2)^{p-2}(1+x^2[2p-1])}
\over{1-\kappa_0(1+x^2)^{p-1}}}.
\label{LW1}
\end{equation}
Comparing this expression for $0<\kappa_0<1$ to the
corresponding expression for $\kappa_0=1$, one can show that the
latter is always greater or equal to the former, 
for any value of $x$, and $p$. Therefore the maximum
value of the image distortion in subcritical lenses of 
Equation~\ref{mass_dist1} can be obtained from Equation~\ref{LW1}
with $\kappa_0=1$ and small $x$. This maximum value, $L/W=3$,
is independent of $\mu$ and $p$. {\it Thus the single image of any 
subcritical lens is rather undistorted, regardless of its impact 
parameter and magnification.}

From Figure~\ref{LWvsAmp1} ~it appears that supercritical
clusters with $\kappa_0\simlt 2$ can produce undistorted images with
$\mu\simgt 10$. We will now show that in a realistic supercritical
cluster the probability of detecting undistorted images is negligible.
To address this, we need to reintroduce the projected source position, 
$y$.  The bottom
panel of Figure~\ref{LWvsx1} shows the distortion and image position
as a function of source position for images of a $\kappa_0=1.1$ lens
(solid lines). It may seem that images II and III are
undistorted. Image III tends to fall very close to the centre of the
cluster, as is seen in the upper panel. Since a core radius,
i.e. $x=1$, typically corresponds to $\sim 50$\h1~kpc, an image at
$x\simlt 0.1$ will probably be hidden within the central cluster
galaxy.  That leaves us with a small range of impact parameters
in the range $\log(y)\sim$ -2 to -1.8 which could produce
detectable undistorted magnified images of type II. For typical cluster
parameters that range translates to roughly 0.1 arcsecond
and is thus smaller than a typical source, i.e. a galaxy several
kpc across at $z_s\sim 0.5-2$.  Additionally, as the $L/W$ ratio
changes very rapidly at these $y$'s, an extended image will appear
distorted.

\subsection{NFW model}\label{NFWmodel}

The NFW lens model is singular at the centre for all values of
$\kappa_s$ (Equations~\ref{mass_dist_proj2a}, \ref{mass_dist_proj2b}),
and so is formally supercritical. This profile will always produce
three images if the source impact parameter is sufficiently small.

\begin{figure}
\vbox{
\centerline{
\psfig{figure=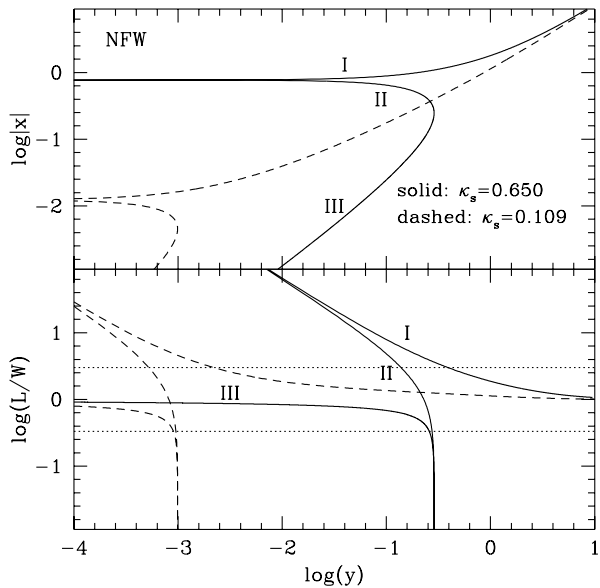,width=3.25in,angle=0}
}
\caption[]{Image position, $\vert x\vert$, (top panel) and 
length-to-width ratio 
(bottom panel) vs. source impact parameter, $y$, for two lenses:
$\kappa_s=0.650$ (solid lines) and $\kappa_s=0.109$ (dashed
line) of the NFW cluster model. Roman numerals indicate image type.
\label{LWvsx2}
}
}
\end{figure}

\begin{figure}
\vbox{
\centerline{
\psfig{figure=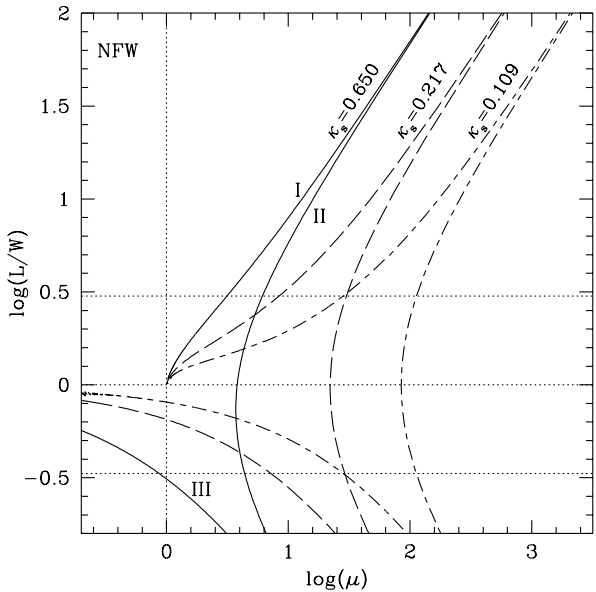,width=3.25in,angle=0}
}
\caption[]{Magnification vs. distortion of images formed by an NFW 
cluster model, represented by Equations~\ref{mass_dist_proj2a},
\ref{mass_dist_proj2b}.
All clusters, regardless of $\kappa_s$ have three images each if
source impact parameter is sufficiently small.
The images are labeled as in Figure~\ref{LWvsx2}.
Horizontal dotted lines at $L/W=3$ and ${1\over 3}$ mark 
the region of undistorted images. The three values of $\kappa_s$ 
represented in this Figure, 0.650, 0.217, and 0.109,
were picked such that these clusters have the same
mass inside 2.5\h1~Mpc as the ISC ones with $\kappa_0$=3.0, 1.0, and
0.5, respectively.
\label{LWvsAmp2}
}
}
\end{figure}

Figure~\ref{LWvsx2} shows the relation between the image position,
$x$ and source position, $y$. Note that
the  values of $\kappa_s$ were picked such that the total mass
within 2.5\h1~Mpc of these clusters is the same as that of ISC
clusters with $\kappa_0$=1.1, and 0.5 (see Section~\ref{clusters}).
The images are labeled as in Figures~\ref{LWvsx1} and \ref{LWvsAmp1},
and have similar properties to those described in
Section~\ref{ISCmodel}. The major difference when compared to the ISC
model is that there is no NFW case that would correspond to a
subcritical, i.e. single-image-only ISC cluster. In fact, relations
between image properties, $y$ vs. $x$ (Figure~\ref{LWvsx2}), and $L/W$
vs. $\mu$ (Figure~\ref{LWvsAmp2}), are qualitatively the same
regardless of the value of $\kappa_s$. In particular, the distortion 
of the primary images of all NFW clusters is an increasing function of 
magnification; therefore giant arcs would be a common type of image in 
these clusters.

Clusters with $\kappa_s\simgt$0.3 cannot have magnified undistorted
images at all (Figure~\ref{LWvsAmp2}), but those with smaller values
of $\kappa_s$ have parts of all three image branches in the HMU
region.  Since we are interested in HMUs let us derive equations
describing image properties for small $\kappa_s$ cases. We will take
advantage of the fact that for small $\kappa_s$ models the triple
image region moves towards small $x$, see top panel of
Figure~\ref{LWvsx2}.  Equation (2.10) of Bartelmann (1996) gives
function $g(x)$, which is proportional to the mass interior to $x$;
$m(x)=4\kappa_s g(x)$.
For small $x$, this simplifies to
\begin{equation} 
g(x)\approx -({\frac{x}{2}})^2(2~{\rm ln}[{\frac{x}{2}}]+1),
\label{gx2}
\end{equation}
and the lens equation~\ref{lens_eq} becomes,
\begin{equation}
y\approx x(1+2\kappa_s{\rm ln}[{\frac{x}{2}}]+\kappa_s).
\label{lens_eq2}
\end{equation}
The condition for radial critical curve is $dy/dx=0$, which can be 
solved for the radius of the corresponding radial caustic in the 
source plane,
\begin{equation}
y_r\approx -4\kappa_s~e^{-(3/2+1/[2\kappa_s])}.
\label{yr2}
\end{equation}
This equation shows that no matter how small $\kappa_s$, $y_r$ will
remain finite, and therefore all NFW clusters can produce 3 images.

The corresponding expressions for magnification and length-to-width of
the images are derived from Equations~\ref{lens_eq},
\ref{mass_dist_proj2a}, \ref{mass_dist_proj2b}, and \ref{mass_dist2},
assuming small $x$:
\begin{equation}
\mu\approx \Big\vert(1+3\kappa_s+2\kappa_s {\rm ln}[{\frac{x}{2}}])
(1+\kappa_s+2\kappa_s {\rm ln}[{\frac{x}{2}}])\Big\vert^{-1}
\label{amplif2}
\end{equation}
\begin{equation}
L/W\approx \Big\vert{\frac{(1+3\kappa_s+2\kappa_s {\rm ln}[{\frac{x}{2}}])}
{(1+\kappa_s+2\kappa_s {\rm ln}[{\frac{x}{2}}])}}\Big\vert
\label{LW2}
\end{equation}
These demonstrate that a line of constant $L/W$ in
Figure~\ref{LWvsAmp2} intersects the primary image branches of various
$\kappa_s$ models at magnifications such that
$\mu\propto \kappa_s^{-2}$. Therefore as one goes to less massive clusters
the magnification of primary images of a given fixed distortion
increases. Therefore highly magnified undistorted images of 
infinitesimally small sources are possible with low-mass NFW clusters.

To determine if these images will stay undistorted for sources of
finite size, one needs to look at how $L/W$ changes with $y$, i.e. the
bottom panel of Figure~\ref{LWvsx2}. For $\kappa_s=0.109$ images II
and III arise only for sources at $y\simlt$0.001. For typical $r_s$ of
300\h1~kpc, this corresponds to $\sim 0.1$ arcsecond, which is smaller
than the expected source size. Therefore images of type II and III
would not produce HMUs.  The primary image shows more promise; the
region of $\mu\simgt$10 in Figure~\ref{LWvsAmp2} corresponds to source
position range $y$ $\sim$ few 0.001-- 0.01, or $\sim$ 1 arcsecond.

\section{Model Assumptions}\label{assumptions}

In the last Section we have shown that both ISC and NFW models can produce
HMUs under certain conditions. Next, we need to estimate the relative
frequency of occurrence of HMUs with the two cluster models.  Assuming
that the cluster and the sources are at fixed redshifts, this involves
two steps: first, for a given cluster one has to integrate over source
impact parameters, taking into account finite source size, and second,
integrate over the distribution of clusters properties.

\subsection{Galaxy clusters}\label{clusters}

We assume that galaxy clusters form a one-parameter family, with the
core radius $r_c$ of ISC and scale radius $r_s$ of NFW being fixed,
but possessing a range of masses.  The cluster mass function, $dN(M)/dM$,
is derived from two observed relations of X-ray selected clusters.
Based on a sample of EMSS clusters, Henry et al. (1992) derive cluster
luminosity function to be, $dN/dL_X \propto L_X^{-\alpha_X}$, where
$\alpha_X$ varies with redshift but is approximately 0.3 between the
redshifts of 0.15 and 0.6.  A relation between cluster bolometric
luminosity, derived from EXOSAT data, and velocity dispersion is 
given by Edge \& Stewart (1991): $L_X \propto \sigma_v^{-2.9}$. 
Combining these two relations
with the assumption that $M \propto \sigma_v^2$, we arrive at cluster
mass function,
\begin{equation}
dN(M)/dM\propto M^{-3.9}.
\label{MF}
\end{equation} 
The absolute normalization is
irrelevant, since we will only be dealing with ratios.  We need not
assume an upper or lower cluster mass limits: the lower mass cutoff is
effectively achieved because low mass clusters have a negligible
lensing cross-section, while at the upper mass end the lensing
cross-section increases slower than the rate at which the numbers of
clusters decrease due to the steep slope of their mass function.

We fix the core radius $r_c$ at 50\h1~kpc, which is substantially
smaller than the derived X-ray `core' sizes, but is consistent with
lensing observations (Fort \& Mellier 1994). For the NFW clusters,
we fix $r_s$ at 300\h1~kpc, which is close to a typical value
obtained in the Navarro et al. (1996) simulations.

With the physical length scale of the two cluster models fixed, we can
now derive the correspondence between the one-parameter ISC and NFW
models, i.e. we ask what is the relation between $\kappa_0$ and
$\kappa_s$ of clusters that have the same mass within 2.5\h1~Mpc. 
The latter is roughly equal to $r_{200}$, and corresponds to 50$r_c$ and
8.33$r_s$, respectively. Using Equations~\ref{mass_dist_proj1},
\ref{mass_dist_proj2a} and \ref{mass_dist_proj2b} we obtain
$\kappa_0\approx 4.615\kappa_s$.

\subsection{Sources}\label{sources}

We assume that the unlensed parent population of HMUs is the same
as that of arcs in clusters. The half-light radii of the sources
of arc images are almost the same as the observed widths of the arcs, 
because the cluster potential is not expected to distort tangential 
arc images in the radial direction. Smail et al. (1996) measure 
half-light radii for a sample of 8 HST arcs (see their Figure 5).
The average half-light radius is about 0.5 arcseconds, which is what
we will adopt in the present paper. We further assume that all the 
sources are circular with a uniform surface brightness.

At any given redshift
the luminosity function (LF) of sources is assumed to be a power law, 
with the slope corresponding to that of the Schechter LF, $\alpha$. 
The value of $\alpha$ is estimated to be about 1.1\quad locally, but may
have been steeper in the past, $\alpha\sim 1.5$ (Ellis et al. 1995). 
We use both values below, to account for the possible range of
$\alpha$'s depending on the type of galaxies and their evolution. 
The results are not very sensitive to $\alpha$. We neglect the sources 
brighter than $L_\star$, the characteristic luminosity of Schechter LF,
because of their small numbers.

We need not make any further assumptions about the source luminosity
function, as we explain below. The number of images of type i, where i 
can be HMU or arcs, for a given cluster characterized by $\kappa_0$ or 
$\kappa_s$ (or cluster mass $M$), is given by,
\begin{equation}
n_{\rm i}\propto \int y~dy~\int_{L_\star}^{L_{lim}/\mu(y)}
\Bigr({L\over L_\star}\Bigl)^{-\alpha}dL \cdot H_{\rm i}
\end{equation}
where $H_{\rm i}$ is the Heviside step function which is 1 if the
image selection criteria are satisfied (Section~\ref{criteria}), 
and 0 otherwise; $L_{lim}$ is the faintest observable luminosity.
The outer integral is over the source impact parameter, 
$y$ in the source plane. The integral can be written as
$$
n_{\rm i}\propto \int y~\Bigr[\Bigr({L_{lim}\over\mu(y)L_\star}\Bigl)
^{1-\alpha}-1 \Bigl]~dy \cdot H_{\rm i}  {\hskip 2in}
$$
\begin{equation}
{\hskip 0.12in} \approx
\Bigr({L_{lim}\over L\star}\Bigl)^{1-\alpha}\int y~[\mu(y)]^{\alpha-1}dy
\cdot H_{\rm i},
\label{n}
\end{equation}
The function $\mu(y)$ is determined by the ISC or NFW model.
The last step in the above equation is justified because the minimum
magnification $\mu$ required for a detectable image is large
(Section~\ref{criteria}), and $\alpha$ is always $>$1. 
Both $L_{lim}$ and $L\star$ are functions of source redshift, but
since we are only interested in the ratio of $n_{\rm HMU}$ to 
$n_{\rm Arcs}$, the dependency on these quantities cancels out.

As the results are quite insensitive to $z_s$ we make no
assumptions about the source redshift distribution.

\subsection{Image selection criteria}\label{criteria}

Images are selected based on their size or morphology. To be selected
as either a giant arc or a HMU, an image has to be lensed appreciably. 
For an arc to be detected, its $L/W$ has to exceed 10, a commonly
used criterium for giant luminous arcs.  A HMU is defined as an image
with central $\mu>10$, and $L/W<3$.  Additionally, the undistorted
nature of the image is guaranteed by requiring that the change in
$L/W$ ratio across the image should not exceed 50\%.

\begin{figure}
\vbox{
\centerline{
\psfig{figure=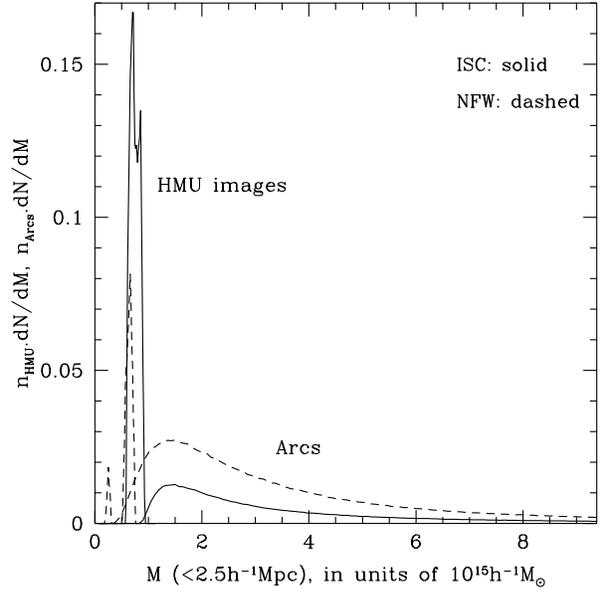,width=3.25in,angle=0}
}
\caption[]{Number of HMU and arc images convolved with the cluster
mass function, as a function of cluster mass. The vertical axis has
arbitrary normalization. Solid lines are for the ISC model, while
dashed lines are for the NFW model. 
Source and lens redshifts were fixed at $z_s=0.3$ and $z_l=1.0$;
Schechter LF slope was assumed to be $\alpha=1.5$.
\label{hmus}
}
}
\end{figure}

{
\begin{table}
\begin{center}
\begin{tabular}{|c|c|c|c|}
\hline
      &             & $z_s$       &             \\ 
$z_l$ & 0.5         & 1.0         & 2.5         \\ \hline
0.2   & 1.28 (1.53) & 1.22 (1.46) & 1.21 (1.44) \\
0.3   & 1.27 (1.55) & 1.16 (1.40) & 1.14 (1.37) \\
0.4   & 1.42 (1.78) & 1.13 (1.36) & 1.08 (1.31) \\ \hline
\end{tabular}
\caption[]{The ratio of the number of HMU images to arcs in the ISC 
model, i.e. $f_{\rm ISC}$, for a range of source, $z_s$ and lens, $z_l$ 
redshifts. Two values of the Schechter LF slope $\alpha$ were tried:
1.5, and 1.1 (in parenthesis).}
\label{table1}
\end{center}
\end{table}

\begin{table}
\begin{center}
\begin{tabular}{|c|c|c|c|}
\hline
      &             & $z_s$       &             \\ 
$z_l$ & 0.5         & 1.0         &  2.5        \\ \hline
0.2   & 0.35 (0.47) & 0.34 (0.44) & 0.33 (0.44) \\
0.3   & 0.25 (0.34) & 0.23 (0.30) & 0.19 (0.30) \\
0.4   & 0.20 (0.29) & 0.16 (0.22) & 0.16 (0.21) \\ \hline
\end{tabular}
\caption[]{Same as Table~\ref{table1}, but for the NFW model.}
\label{table2}
\end{center}
\end{table}
}

\section{Results}\label{results}

Figure~\ref{hmus} shows the numbers of giant arcs and HMUs per cluster,
weighted by the cluster mass function. The numbers plotted along
the vertical axis are proportional to $n_{\rm i}\cdot dN(M)/dM$, where
$n_{\rm i}$ is given by Equation~\ref{n}, and $dN(M)/dM$ by
Equation~\ref{MF}. The solid and dashed lines represent ISC and NFW 
models, 
respectively, and arcs and HMUs are labeled. Here we assumed $z_l=0.3$, 
$z_s=1.0$, and $\alpha=1.5$. Notice that for both mass profiles, HMU 
images are produced by low mass clusters only, as was explained in 
Sections~\ref{ISCmodel} and \ref{NFWmodel}. 
The relative numbers of HMU and arc images produced by ISC and
NFW profiles can be understood intuitively. The shapes of images
depend on the mass density gradient at the location of the images
(Section~\ref{models}). The core region of ISC is flat, hence it
produces roughly equal radial and tangential magnifications,
and is therefore ideal for generating HMUs. 
The projected density of the NFW profile, on the other hand, goes as
-[ln${x\over 2}$+1] at small $x$, and so does not have a flat core.
It is also steeper than isothermal at large radii. Overall, the
NFW profile is steeper than ISC, and thus is better than ISC at making 
giant arcs, and worse than ISC at making HMUs.

We are interested in the ratio of HMUs to giant arcs for a given cluster
model, i.e.,
\begin{equation}
f_{\rm model}=
{\int n_{\rm HMU}(M)\cdot [dN(M)/dM]~dM\over
\int n_{\rm Arcs}(M)\cdot [dN(M)/dM]~dM},
\label{f}
\end{equation}
where the model is either ISC or NFW. This is just the ratio of
the areas under the HMU and arc curves in Figure~\ref{hmus}.
For the parameters of the Figure, $f_{\rm ISC}=1.16$, and 
$f_{\rm NFW}=0.23$. 

Table~\ref{table1} shows how the $f_{\rm ISC}$ ratio changes with 
source and lens redshift. The numbers (in parentheses)
are for the Schechter LF slope $\alpha$ of 1.5 (1.1). 
Table~\ref{table2} contains the corresponding values for the
NFW model. It is seen from these two tables that the predictions are 
virtually independent of source redshift, but depend on lens 
redshift. The source and lens redshift come in only in $\Sigma_{crit}$
(Equation~\ref{critical}),
and both of these trends arise because of the way $\Sigma_{crit}$,
which determines cluster's lensing strength, varies with $z_l$ and $z_s$. 
When $z_l$ is held constant, $\Sigma_{crit}$ attains its asymptotic 
value at redshifts just beyond $z_l$, and thus source redshift has 
very little effect on $f$. On the other hand, when $z_s$ is 
constant, $\Sigma_{crit}$ changes quite rapidly at low--moderate
lens redshifts. The sense of the trend is also understood in terms
of $\Sigma_{crit}$: when the lensing strength of a cluster of a certain
mass increases due to the decrease in $\Sigma_{crit}$, more arcs are
produced, and hence $f$ goes down. The weak dependency on $z_s$ is a 
useful feature because source redshift distribution is arguably the 
most uncertain of the relevant model parameters. 

The $f$ ratios depend somewhat on the source LF slope, $\alpha$. The 
dependency is not very strong because most of the images are magnified 
just by the minimum required $\mu$ (because the cluster cross-section
declines rapidly with $\mu$), and so both HMUs and arcs are drawn from 
a rather narrow portion of the LF. The decline in cluster cross-section 
with $\mu$ is more severe for HMUs because of the additional selection 
restrictions placed on undistorted images. Therefore, when $\alpha$ is
steeper and there are more faint sources, the number of HMU images does 
not increase as much as the number of arcs; hence $f$ is smaller for
steeper LFs.

Dependence on cosmology is weak. Tables~\ref{table1} and \ref{table2} 
assume flat Universe model, dominated by a cosmological constant;
$\Omega=0.2$, $\Lambda=0.8$. Adopting a matter-dominated flat Universe,
$\Omega=1.0$, produces at most a 40\% increase in $f$. Since $f$ is 
the ratio of the numbers of images, it does not depend on the 
Hubble constant.

Note that the predictions for relative 
numbers of HMUs to giant arcs are a function of chosen image selection 
criteria. For example, if minimum HMU magnification is increased from 
10 to 30, then $f_{\rm ISC}$ drops by about a factor or 10 compared to 
the tabulated values, but virtually no HMUs are expected with the 
universal dark matter profile.

\section{Conclusions}\label{conc}

In this paper we considered two cluster mass density profiles,
isothermal sphere with a core and 
a universal dark matter profile, and studied the properties of
gravitationaly lensed images of extented sources produced by
these models. We were particularly interested in highly magnified
undistorted images, and have shown that both profiles can in 
principle produce such images, though when realistic cluster properties
and mass functions were folded in, ISC models proved to be much more
efficient than NFW at making HMU images. Using simplified assumptions,
we calculated $f$, the ratio  of HMUs to giant arcs for the two models, 
as a function of source and lens parameters. To account for cluster
asymmetries and substructure, which were not considered in this paper,
the derived $f$ ratios should be multiplied by the ratio of the frequency 
of arcs in substructured clusters to that in smooth symmetric clusters. 
The latter ratio can be obtained from numerical simulations of the type
described in Bartelmann et al. (1995). We have shown that $f$ 
is not very sensitive to cosmology, source luminosity function and 
redshift distribution, and lens redshift distribution. 

In fact,
the strongest dependency by far is on the cluster model: with the ISC 
model HMUs are on average 1.2 times as abundant as giant arcs, whereas 
with the NFW model HMUs are only 0.2 times as common as arcs (see
Tables~\ref{table1} and \ref{table2}).
{\it Thus relative frequency of highly magnified undistorted images 
to giant arcs can be used to discriminate between isothermal clusters 
with cores and universal dark matter halo profiles},
if a complete sample of clusters is examined for HMU/arc images. 
Two points need to be stressed here.
First, since HMUs are produced by clusters at the lower end of the mass 
function, $M(<2.5h^{-1}{\rm Mpc})~\simlt~10^{15}h^{-1}~M_\odot$, a cluster 
sample should extend down to such masses, or corresponding X-ray 
luminosities; alternatively, model predictions should make an allowance 
for a high-mass cluster cutoff.
Second, HMUs would not be as easily detected as giant arcs, since
they would have regular morphology. If they are at high redshift their 
surface brightness will be correspondingly low. However, they should
be appreciably larger than typical high-$z$ galaxies, with half-light 
radii of about 1.5--2 arcseconds. 
To find HMUs one would need to do spectroscopy on extended, low
surface brightness galaxies with regular morphology, 
located within 10-20 arcseconds of cluster centres, in clusters
with no giant arcs. Extensive searches with such observational 
constraints have not been undertaken; therefore it is not surprising 
that highly magnified undistorted images, whether their number density 
on the sky is comparable to or much smaller than that of giant arcs, 
have so far escaped detection. 

\section*{Acknowledgments}

We would like to thank Alastair Edge, Richard Ellis, and Mike Irwin
for helpful suggestions regarding an early version of the paper.
LLRW would like to acknowledge the support of PPARC Fellowship
at the Institute of Astronomy, Cambridge, UK.

\end{document}